\documentclass[aps,pra,reprint,superscriptaddress,hidelinks,nobalancelastpage,longbibliography]{revtex4-1}
\usepackage{amsmath,braket,amssymb}
\usepackage{graphicx}
\usepackage{color}
\usepackage{hyperref}

\usepackage[separate-uncertainty=true,multi-part-units=single]{siunitx}

\renewcommand{\vr}{\textbf{r}}
\newcommand{\vE}{\textbf{E}}

\begin{document}
\title{Dark Exciton Preparation in a Quantum Dot by a Longitudinal Light Field Tuned to Higher Exciton States}

\author{M. Holtkemper}
\affiliation{Institut f\"ur Festk\"orpertheorie, Universit\"at M\"unster, Wilhelm-Klemm-Str.~10, 48149 M\"unster, Germany}

\author{G. F. Quinteiro}
\affiliation{IMIT and Departamento de F\'isica FaCENA, Universidad Nacional del Nordeste, Corrientes, Argentina}

\author{D. E. Reiter}
\affiliation{Institut f\"ur Festk\"orpertheorie, Universit\"at M\"unster, Wilhelm-Klemm-Str.~10, 48149 M\"unster, Germany}
\email{Doris.Reiter@uni-muenster.de}

\author{T. Kuhn}
\affiliation{Institut f\"ur Festk\"orpertheorie, Universit\"at M\"unster, Wilhelm-Klemm-Str.~10, 48149 M\"unster, Germany}
\email{Tilmann.Kuhn@uni-muenster.de}

\date{\today}

\begin{abstract}
Several important proposals to use semiconductor quantum dots in quantum information technology rely on the control of the dark exciton ground states, such as dark exciton based qubits with a $\mu$s life time. In this paper, we present an efficient way to occupy the dark exciton ground state by a single short laser pulse.
The scheme is based on an optical excitation with a longitudinal field component featured by, e.g., radially polarized beams or certain Laguerre-Gauss or Bessel beams. Utilizing this component, we show within a configuration interaction approach that high-energy exciton states composed of light-hole excitons and higher dark heavy-hole excitons can be addressed. When the higher exciton relaxes, a dark exciton in its ground state is created.
\end{abstract}

\maketitle

\section{Introduction}\label{sec:introduction}
With their discrete energy states, semiconductor quantum dots (QDs) are designated to play an important role in solid-state quantum information technology \cite{michler2017quantum}, for example as sources for single or entangled photons \cite{gazzano2016toward,senellart2017high,huber2018semiconductor,portalupi2019inas,rodt2020deterministically} or as realization of qubits \cite{biolatti2000quantum,ramsay2010review,warburton2013single}. For qubits, it is especially promising to use the dark excitons as information storage, since they exhibit extraordinary long life and coherence times \cite{mcfarlane2009gigahertz,johansen2010probing,poem2010accessing,huo2017spontaneous,heindel2017accessing}.
Furthermore, dark excitons are important intermediate states in some state preparation protocols, e.g. for biexciton \cite{luker2019review} or impurity spin preparation \cite{reiter2009all} and they have been used for the generation of a cluster state of entangled photons \cite{schwartz2016deterministic}.

The long lifetime of dark excitons is mainly due to the vanishing or very weak dipole coupling to the light field which, in turn, makes their optical generation challenging. Several workarounds have been developed to overcome this difficulty. By using a non-resonant cw excitation and a subsequent random optical charging of the QD one can form excited biexcitons with various spin configurations. After emission of a photon and subsequent phonon-induced hole relaxation certain spin configurations lead to the formation of a dark exciton \cite{poem2010accessing}. Alternatively, weak mixing between bright and dark excitons can be induced either by applying an external magnetic field \cite{luker2015direct,germanis2018dark} or by breaking the $C_{2v}$-symmetry of the QD \cite{zielinski2015atomistic}. This makes a direct optical excitation of the dark exciton possible \cite{schwartz2015deterministic}, however with an oscillator strength which is some orders of magnitude smaller than the absorption into the bright exciton. A (desired) increase of the optical excitation by a stronger coupling between dark and bright excitons is unevitably associated with a further (undesired) distortion of the spin character of the dark exciton resulting, e.g., in a decrease of the lifetime.

In this paper, we propose a scheme to circumvent this connection of absorption strength and spin distortion by moving the necessary spin coupling to higher excited states. There the spin coupling can be strong essentially without affecting the pure spin character of the ground states. In detail, our scheme utilizes suitable complex light fields with a strong longitudinal field component. Such field components can excite otherwise optically forbidden light-hole (LH) excitons. Valence-band mixing couples those LH excitons preferentially to higher heavy-hole (HH) excitons involving a dark spin configuration, building strongly-mixed higher exciton states. Due to their dark HH exciton contribution, these states will then relax by phonon emission into the dark HH exciton ground state. 

Our proposal is related to the findings in Ref.~\onlinecite{smolenski2012plane}, where a small coupling between the dark HH ground state and a LH exciton with total angular momentum zero was found to enable an optical recombination of the dark exciton ground state by emitting a photon in in-plane direction with a polarization in $z$-direction. This recombination was identified as the main relaxation process of the dark ground state. However, the coupling is small, because the lowest LH exciton and the dark HH ground state are strongly separated in energy. Considering higher excited HH-shells, like the $d$-shell, we will show that these couplings become strong and enable the proposed efficient scheme to excite the dark HH ground state. Our proposal thus complements studies on the great benefits of the use of higher QD exciton states \cite{huneke2011role,suzuki2018dephasing,hinz2018charge,holtkemper2018influence, qian2019enhanced} or complex light fields beyond Gaussian beams \cite{quinteiro2015formulation,quinteiro2014light} for QD based quantum information technology.

The paper is structured as follows: In Sec.~\ref{sec:light} we describe light modes which carry strong components of the required polarization in the growth direction. Sec.~\ref{sec:states} then introduces our QD model and basic selection rules for the transverse and longitudinal light modes. The absorption spectra corresponding to these modes are discussed in Sec.~\ref{sec:spectra}. Based on this knowledge, we present the excitation scheme for the dark exciton ground states in Sec.~\ref{sec:scheme}. Some conclusion are given in Sec.~\ref{sec:discussion}.

\section{A light field polarized in the growth direction}\label{sec:light}
A key element for our excitation process is a component of the light field polarized in the growth direction of the QD sample. Such a polarization could be achieved by a laser entering the QD sample from a cleaved edge \cite{huo2014light,huo2017spontaneous}. However, cleaving the sample is often associated with other drawbacks, e.g., by introducing additional surfaces close to the QD. Therefore, it is desirable to use a beam at normal incidence along the growth direction, which has a component of the electric field polarized along the propagation direction. In the following we will call this a longitudinal component. While such a longitudinal field component is absent in plane wave-like fields and is very small in beams which are well described by the paraxial approximation, it can become dominant close to the beam axis in certain strongly non-paraxial or tightly focused beams, such as certain Laguerre-Gaussian, Bessel or radially polarized beams \cite{youngworth2000focusing,novotny2001longitudinal,zhan2004trapping,quinteiro2017twisted}. This dominance can be strong enough to safely neglect all but the longitudinal field components in the region close to the beam center. Such longitudinal components have been successfully used to probe single molecules \cite{novotny2001longitudinal} or to trap metallic Rayleigh particles \cite{zhan2004trapping}. 

To provide a clear and well-defined theoretical description, here we consider Bessel beams, which constitute an exact and complete set of solutions to Maxwell's equations. They are propagation-invariant (non-diffracting) beams. Although strictly speaking Bessel beams have an infinite lateral extension because of their weak decay in the radial direction, approximations of such beams have been realized in various experiments  \cite{durnin1987diffractionfree,woerdemann2013advanced,ettorre2015experimental, milione2015measuring}. Furthermore, it can be shown that the tight focusing of cylindrically symmetric vector beams typically also results in Bessel function-like field distributions in the transverse and longitudinal components \cite{youngworth2000focusing}.

Assuming a beam propagating along the $z$-direction with $\vE(\vr,t)=\frac{1}{2} \tilde{\vE}(\vr)e^{i(q_z z-\omega t)} + \textrm{c.c.}$, where $\textrm{c.c.}$ denotes the complex conjugate, the electric field of a Bessel beam reads
\begin{subequations}
\label{Eq:E_Bessel}
\begin{eqnarray}
\label{Eq:Exy}
    \tilde{E}_x(\mathbf r)  &=& i\sigma  \tilde{E}_y(\mathbf r) =   i  \frac{E_0}{\sqrt{2}}  J_{\ell}(q_r r) e^{i \ell \varphi}\\
\label{Eq:Ez}
    \tilde{E}_z(\mathbf r) &=&   \sigma \frac{E_0}{\sqrt{2}}  \frac{q_r}{q_z}  J_{\ell+\sigma}(q_r r)    e^{i (\ell+\sigma) \varphi}   \hspace{7 mm}
\end{eqnarray}
\end{subequations}
with the Bessel function of first kind and $\ell$-th order $J_{\ell}$, the electric field amplitude $E_0$, the frequency $\omega$ and the wave vector components in propagation and radial directions, $q_z$ and $q_r$, respectively. The latter quantities are related by $q_z^2+q_r^2 = (n\omega/c)^2$, $n$ being the index of refraction of the medium. The integers $\sigma=\pm1$ and $\ell=0,\pm1,...\,$ characterize the handedness of circular polarization and the orbital angular momentum of the transverse components of the beam, respectively. We consider a CdSe QD with $n=2.8$ \cite{ninomiya1995optical} and realistic highly non-paraxial (tightly focused) beams with $q_r/q_z=1$ \cite{chen2009realization,huang2015efficient}.

Since for all Bessel functions $J_n$ with $n>0$ the intensity of the beam profile has a minimum around the beam axis (thus at the position of the QD), there are only two types of beams with a high intensity at the QD:
First, beams with $\ell=0$ and $\sigma=\pm1$, which are characterized by a predominantly transverse field component. Selection rules and spectra of these beams are similar to those of typical Gaussian (or plane wave-like) beams. Second, beams with $\ell=\pm1$ and $\sigma=\mp1$, which are characterized by a predominantly longitudinal field component. We note, that one can also superpose the two Bessel beam modes $\ell=\pm1$ and $\sigma=\mp1$ to a radially or azimuthally polarized beam, where the radially polarized beam also features the pronounced longitudinal field component \cite{youngworth2000focusing,novotny2001longitudinal,zhan2004trapping} while the azimuthally polarized beam remains purely transverse with vanishing intensity at the beam center \cite{youngworth2000focusing}. In the following we will consider the absorption by both the transverse and the longitudinal field components.

\section{Electronic states and optical selection rules}\label{sec:states}
\begin{figure}[tb]
\includegraphics[width=0.8\columnwidth]{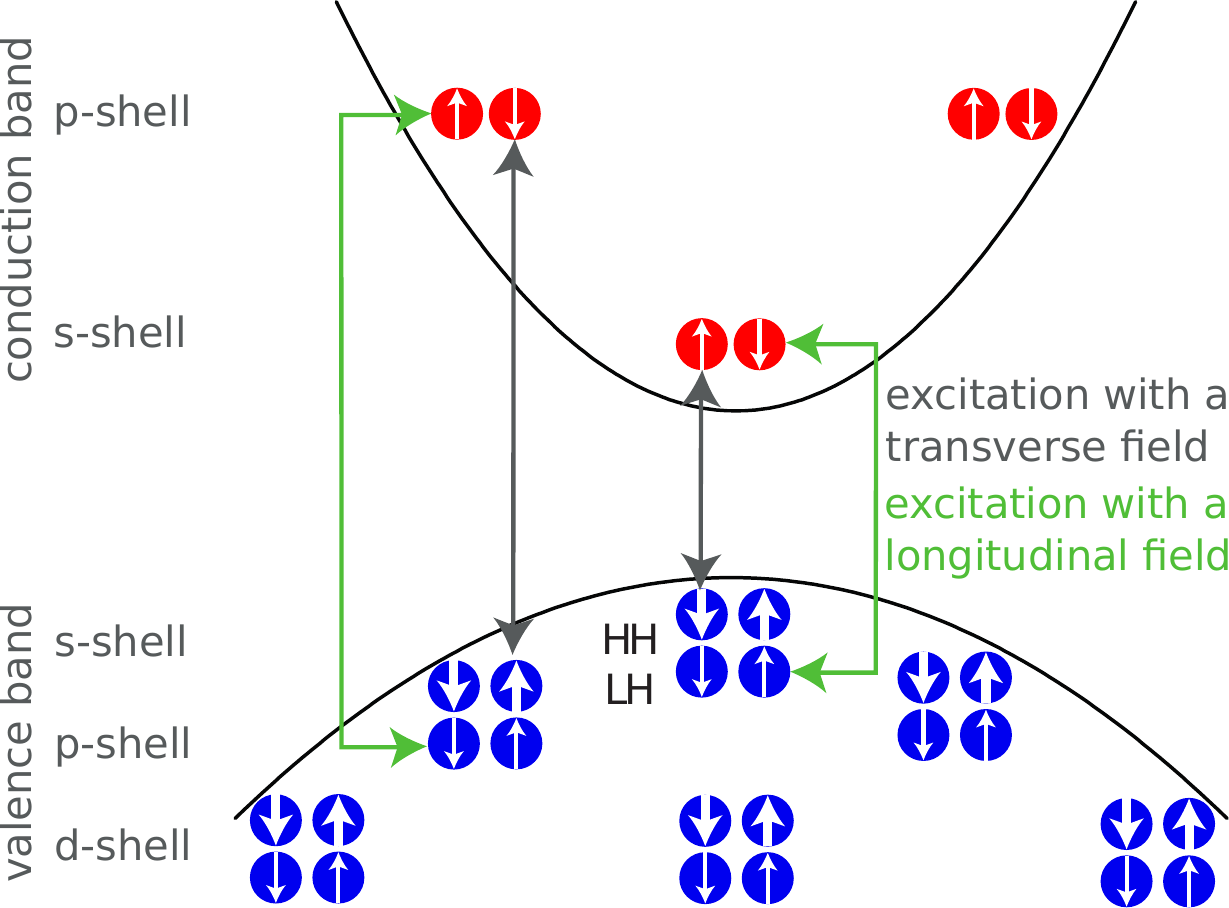}
\caption{
	\label{fig:sketch}
	Sketch of the optical selection rules in a QD. The red dots mark the conduction band electrons and the blue dots represent the valence band holes, while the spin is indicated by a white arrow (bold for HHs and thin for electrons and LHs). The states are grouped in the usual shells $s$, $p$, $d$, ... . The dark gray (light green) arrows mark exemplary electron-hole pairs which can be excited by a transverse (longitudinal) field component.
	}
\end{figure}

We are interested in the optical excitation of a flat self-assembled QD with a diameter of a few nanometers. The confinement of electron and hole states is modeled by a three-dimensional, anisotropic harmonic oscillator potential with confinement lengths $L_x$, $L_y$ and $L_z$. The electronic states are described within an envelope function formalism. Therein, the wave function is separated into a Bloch part and an envelope part. For the Bloch states, we take into account the HH valence band with a total angular momentum projection (or pseudo spin) of $J_h=\pm3/2$, the LH valence band with $J_h=\pm1/2$ and the conduction band for electrons with $J_e=\pm1/2$. According to the confinement, the envelope states are given by harmonic-oscillator basis functions with their lowest-lying states just increasing in the in-plane quantum numbers. As usual \cite{holtkemper2018influence,hawrylak2000excitonic,trojnar2011quantum,smolenski2016fine}, we group the states into $s,p,d,$\ldots ($S,P,D,$\ldots )-shells. We will use small letters for HHs and capital letters for LHs. Excitons as well as their exciting transitions are labeled by the valence band to conduction band state involved (e.g. $d\to s$ is for the excitation of a HH from the $d$-shell to an electron in the $s$-shell).
The Bloch part of the wave function determines the spin selection rules. For transverse electric fields only electron-hole pairs with a total angular momentum projection of $\pm1$ can be excited, i.e., HH$\pm1$ excitons with $J_h=\pm 3/2$ and $J_e=\mp 1/2$ and LH$\pm1$ excitons with $J_h=\pm 1/2$ and $J_e=\pm 1/2$. Transitions into the ``dark'' HH$\pm2$ excitons with $J_h=\pm 3/2$ and $J_e=\pm 1/2$ and LH$\pm0$ excitons with $J_h=\pm 1/2$ and $J_e=\mp 1/2$ are forbidden \cite{poem2010accessing}.
For longitudinal field components, new spin selection rules apply and the light-hole excitons LH$\pm0$ \cite{quinteiro2014light} can be excited, while HH$\pm1$, HH$\pm2$ and LH$\pm1$ are spin forbidden.
The transitions between different shells are governed by the overlap integral of the envelope functions. Since the confinement lengths for conduction and valence band states are taken to be the same, envelope functions from different shells are orthogonal and only transitions between the envelope states from the same shell are possible.
Fig.~\ref{fig:sketch} provides a sketch of exemplary transitions addressable by the transverse (gray arrows) and longitudinal field (light green arrows). The electron and hole states are depicted as red and blue dots, respectively. The white arrows (bold for HHs and thin for electrons and LHs) indicate the different spin states, while the groups of two/four states represent the different envelope states.

\section{Absorption spectra for the transverse and longitudinal modes}\label{sec:spectra}
\begin{figure}[tb]
\includegraphics[width=\columnwidth]{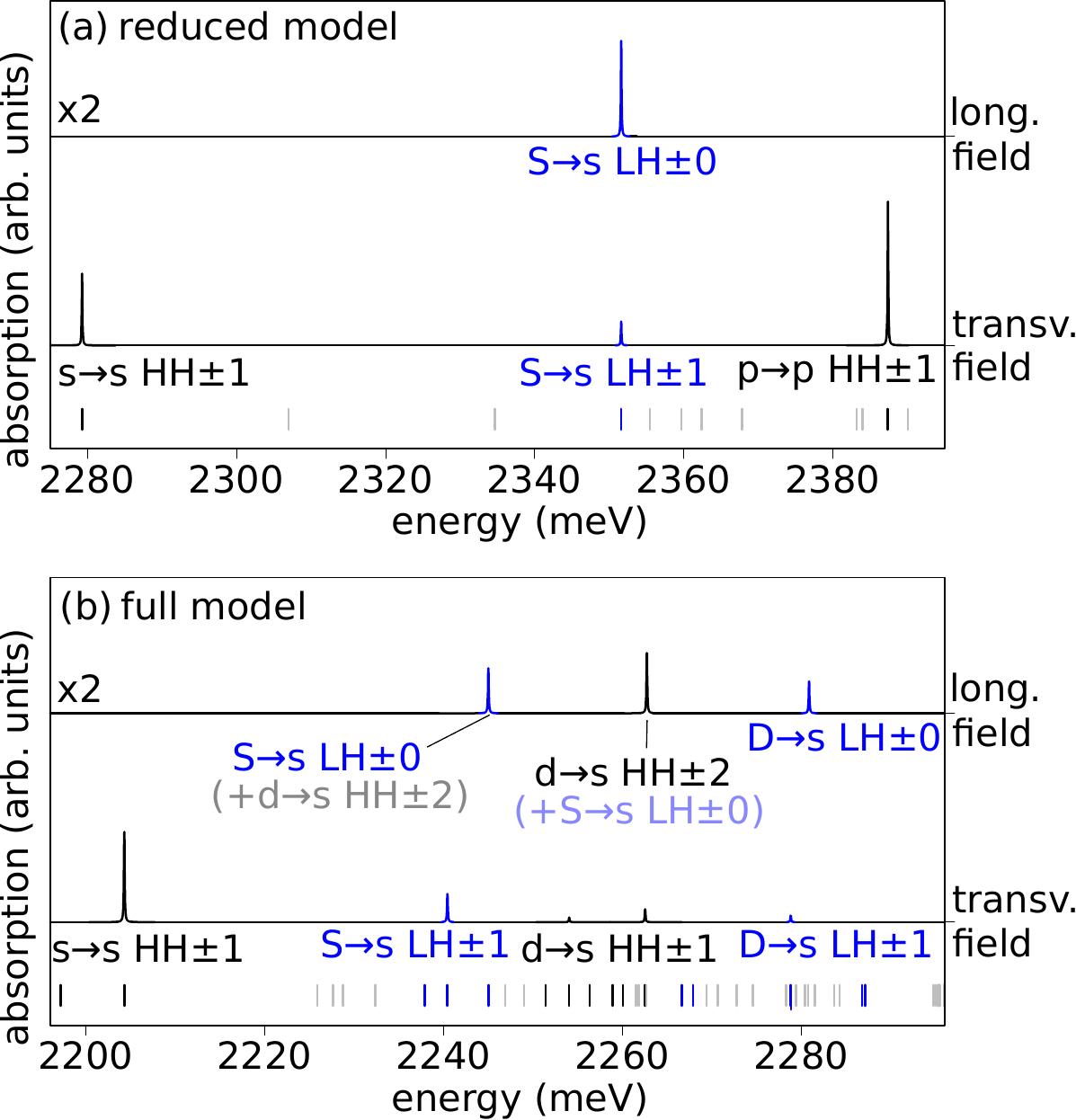}
\caption{
	\label{fig:spectra}
	Optical spectra of a QD in (a) the reduced QD model and (b) the full QD model including Coulomb interaction and valence-band mixing. In each part the bottom line shows the spectrum for the transverse field and the upper line for the longitudinal field. The small lines at the bottom mark the positions of all available states, which in (a) are in general multiply degenerate, while in (b) many degeneracies are lifted.
	}
\end{figure}

Using a simple model of uncoupled electron-hole pairs inside the QD, we can calculate the corresponding absorption spectrum, which is shown in Fig.~\ref{fig:spectra}(a) for the two types of exciting beams. Here we have considered a QD with a size of $(L_x\times L_y \times L_z) = (5.4\times 5.4 \times 2.0)$~nm$^3$ and material parameters for CdSe as in Ref.~\onlinecite{holtkemper2018influence}. The small lines at the bottom mark the positions of all states, which are in general multiply degenerate. 
We find that the corresponding spectra indeed match our expectations. More specifically, for the transverse field (bottom line) we predict the $s\to s$ and $p\to p$ HH$\pm1$ transitions and the $S\to s$ LH$\pm1$ transition. For the longitudinal field (top line) only the $S\to s$ LH$\pm0$ transition is excited in the given energetic range, however, the oscillator strength is similar to values expected for typical transverse field excitations.

\begin{figure}[tb]
\includegraphics[width=1.0\columnwidth]{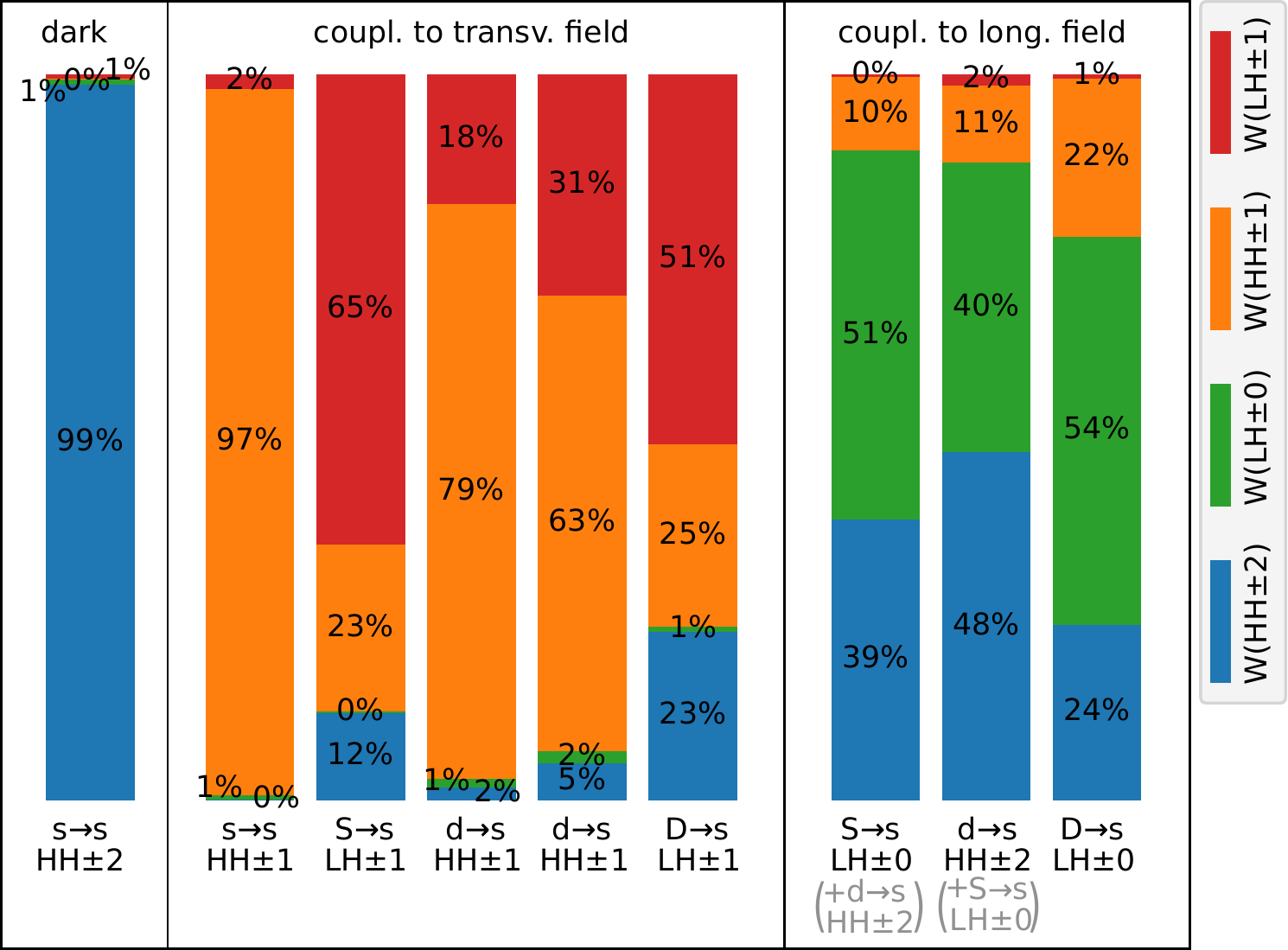}
\caption{
	\label{fig:anteileplot}
	Different spin contributions of the dark exciton $s\to s$ HH$\pm2$ (not directly accessible by optical excitation) and the excitons excited by the transverse and longitudinal fields for a QD with size $(5.4\times 5.4 \times 2.0)$~nm$^3$.
	}
\end{figure}

Valence-band mixing and correlation effects induced by Coulomb interaction become important especially for higher exciton states \cite{holtkemper2018influence}. To get a more realistic description of a QD, we extend the envelope function approximation by including the direct Coulomb interaction, the short range exchange Coulomb interaction as well as valence-band mixing via the Luttinger Hamiltonian within a configuration interaction approach to obtain the exciton eigenstates of the QD. The interactions lead to a strong mixing of the electron-hole pair states. Details of the model, including its limitations, can be found in Ref.~\onlinecite{holtkemper2018influence}.

The corresponding spectra are shown in Fig.~\ref{fig:spectra}(b) for the two types of excitation discussed above. Below the spectra again all existing states are indicated. Many degeneracies seen in the uncoupled model, associated typically with different spin configurations, are now lifted. For example, at the ground state transition ($s\to s$) a splitting into two levels appears, the lower one reflecting the two dark HH$\pm2$ and the upper one the two bright HH$\pm1$ excitons. Both doublets can be further split by breaking the cylindrical symmetry of the QD confinement. The lowest transition involving light-holes, $S\to s$ splits into three levels, a doubly degenerate LH$\pm1$ transition as well as two energetically separated states containing the LH$\pm0$ excitons, from which one is accessible by the longitudinal field.

Further considerations of the spectra reveal that all the transitions from the reduced model still prevail in the full model, however at different energies due to the Coulomb shifts. Also, the relative intensities of the lines change. Note that for the sake of clarity we use the same labels for the transitions. However, due to the mixing there are in general contributions from other electron-hole pair states. In addition, due to the state mixing several new absorption lines show up. In the considered energetic range these are $d\to s$ and $D\to s$ transitions.

It is instructive to understand the spin contributions of the different excitons. We define these spin contributions $W(\text{HH}\pm 2)$, $W(\text{HH}\pm 1)$, $W(\text{LH}\pm 0)$ and $W(\text{LH}\pm 1)$ via the trace over the spatial degrees of freedom. The spin contributions are displayed in Fig.~\ref{fig:anteileplot} for the dark ground state exciton and the excitons excited by the transverse and longitudinal fields (see Fig.~\ref{fig:spectra}(b)). We note that while the dark and the bright HH $s\to s$ excitons exhibit almost no mixing, all other excitons have pronounced contributions from at least two different spin configurations. Especially the predominantly LH eigenstates have strong admixtures of energetically nearby HH states. This holds in particular for those excitons excitable by a longitudinal field. For our QD geometry, the $d\to s$ HH excitons are energetically close to the $S\to s$ LH excitons, which leads to a strong mixing of $d\to s$ HH$\pm 2$ and $S\to s$ LH$\pm 0$. This mixing is so strong, that we include the respective admixture in parenthesis within the labeling of the respective eigenstate and call them $S\to s$ LH$\pm0$ (+$d\to s$ HH$\pm2$) and $d\to s$ HH$\pm2$ (+$S\to s$ LH$\pm0$). In the following, we will focus on these two eigenstates.
Please note, that the contributions of the bright spin states do not represent the brightness of the respective exciton since, e.g., envelope selection rules, which are not considered in the discussion of spin contributions, also affect the brightness.

\section{Scheme to excite the dark exciton ground state}\label{sec:scheme}

\begin{figure}[tb]
\includegraphics[width=1.0\columnwidth]{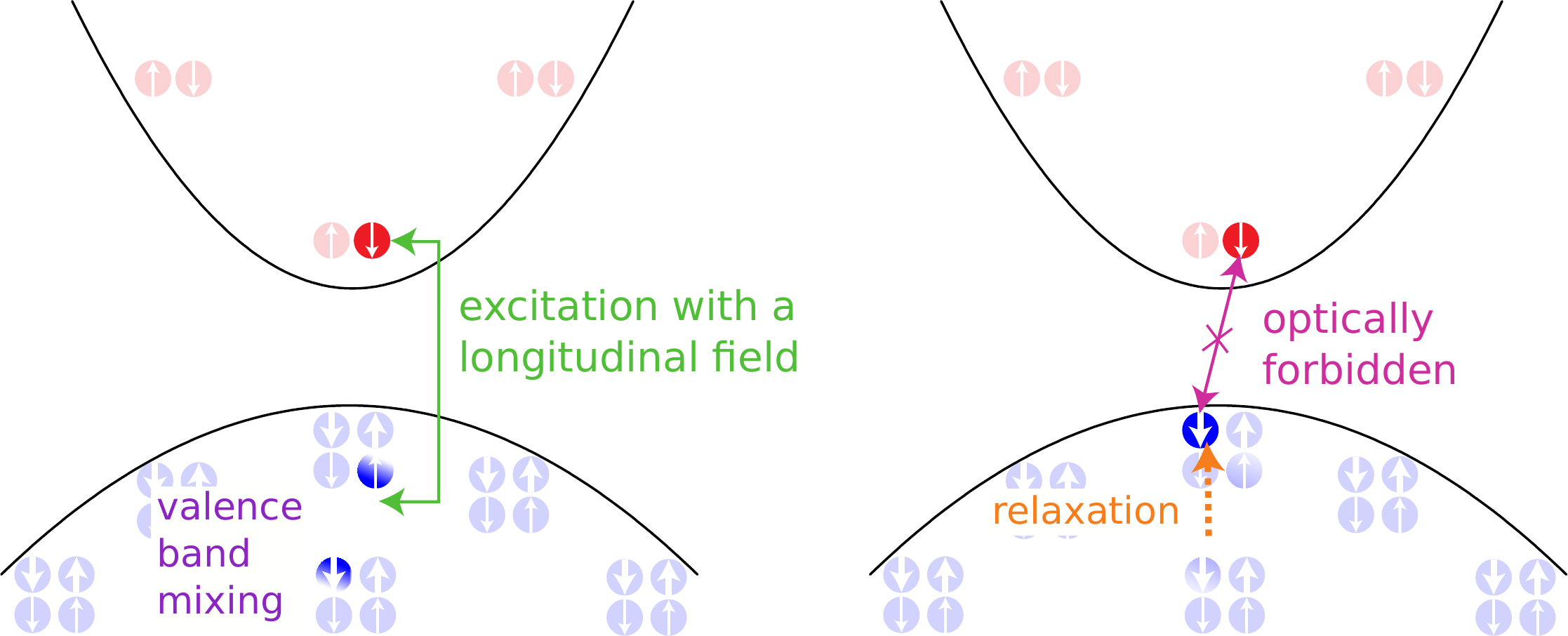}
\caption{
	\label{fig:relax}
Proposed scheme for dark exciton generation. QD states in the valence and conduction band are sketched as in Fig.~\ref{fig:sketch}. On the left, the purple shaded area indicates the valence-band mixing between a $d$-shell HH and a $s$-shell LH. The green arrow describes the optical excitation of a corresponding exciton by the longitudinal component of a light field. On the right, the relaxation of the mixed hole is illustrated by the orange arrow. The optical recombination of the resulting exciton is forbidden (violet arrow).
	}
\end{figure}

The strong spin mixing observed for the states $d\to s$ HH$\pm2$ and $S\to s$ LH$\pm0$ in the full QD model builds the basis for our scheme to excite the dark exciton ground state. The scheme is sketched in Fig.~\ref{fig:relax} and based on two steps:

1. As discussed above, the valence-band mixing causes a coupling between HH$\pm 2$ and LH$\pm 0$, leading to two mixed eigenstates. The LH$\pm 0$ spin components of these eigenstates makes them accessible by the longitudinal light field. We will refer to this optical accessibility via the relative brightness $B$, given in units of the oscillator strength of the unmixed $S \to s$ LH$\pm 0$ exciton. In our case (see Fig~\ref{fig:spectra}), the exciton labeled by $S\to s$ LH$\pm 0$ (+$d\to s$ HH$\pm2$) has a relative brightness of $B=47\%$, while the exciton labeled by $d\to s$ HH$\pm2$ (+$S\to s$ LH$\pm0$) has a relative brightness $B=63\%$.

2. After excitation, higher excitons tend to relax into lower-lying exciton states, a process mediated by the emission of phonons, which takes place typically on a timescale of a few hundred femtoseconds up to some picoseconds \cite{hinz2018charge}, depending on the geometry of the QD. Spin flip processes via the coupling to nuclear spins \cite{merkulov2002electron} occur on a nanosecond timescale and are therefore negligible. Because the emission of phonons does not change spins, the spin state of the initial and final eigenstate of a relaxation step need to be similar. To estimate the probability of different relaxation channels, we compare the spin contributions of all possible intermediate states. If there are no equal spin contributions in two states, a relaxation between these states is forbidden.

\begin{figure}[tb]
\includegraphics[width=1.0\columnwidth]{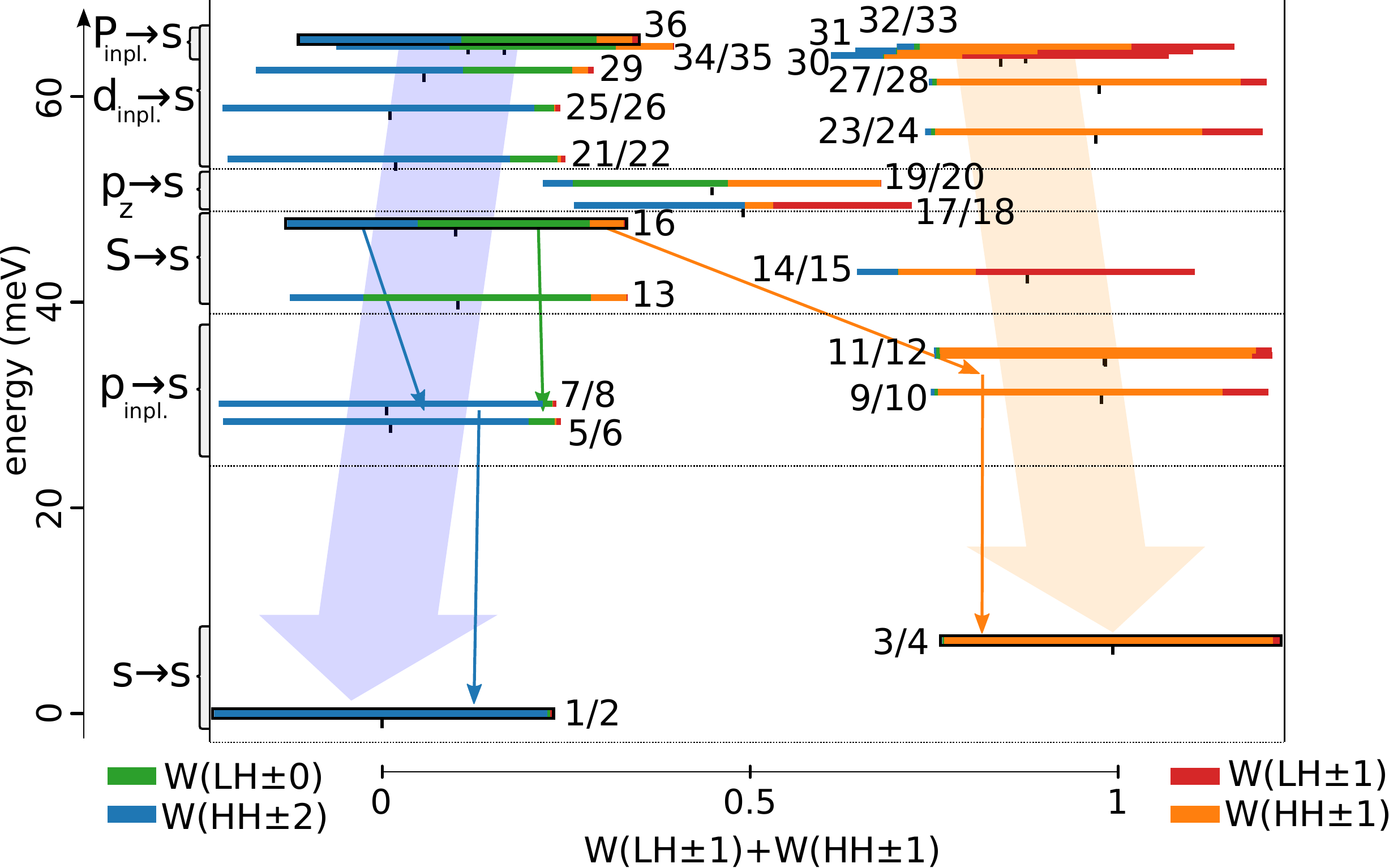}
\caption{
	\label{fig:contribrelaxin}
	Color-bars representing the contributions $W(\text{HH}\pm 2)$, $W(\text{HH}\pm 1)$, $W(\text{LH}\pm 0)$ and $W(\text{LH}\pm 1)$ of the energetically lower exciton states (see small lines below the spectra in Fig.~\ref{fig:spectra}). The color-bars are ordered by energy in vertical direction and by their combined contribution $W(\text{HH}\pm 1) + W(\text{LH}\pm 1)$ in the horizontal direction. A black frame highlights the color-bars representing the two here relevant initial states $S\to s$ LH$\pm0$ (+$d\to s$ HH$\pm2$) and $d\to s$ HH$\pm2$ (+$S\to s$ LH$\pm0$), as well as the ground states. The two shaded arrows in the background mark the general relaxation cascades, while the thinner arrows mark the possible relaxation paths of the $S\to s$ LH$\pm0$ (+$d\to s$ HH$\pm2$) state.
	}
\end{figure}
In Fig.~\ref{fig:contribrelaxin} we show the energy level scheme of the lower exciton up to the $d\to s$ HH$\pm2$ state. The eigenstates are represented by color bars which indicate the respective spin contributions. These color bars are ordered in the vertical direction regarding their energy (measured with respect to the lowest exciton state, i.e., the dark $s\to s$ HH$\pm2$ ground state exciton) and in the horizontal direction regarding their combined contribution $W(\text{HH}\pm 1) + W(\text{LH}\pm 1)$. We observe that this combined contribution essentially separates the states into two spin groups. This separation is caused by the nature of valence band mixing. Regarding the spin contributions, the relaxation of a state into other states of the same spin group is preferred (see broad shaded arrows in the background of Fig.~\ref{fig:contribrelaxin}). However, if the intermediate states have non-vanishing spin contributions of the other spin group, there is a certain chance for relaxations between the spin groups. Focusing on the relaxation of the $S\to s$ LH$\pm 0$ (+$d\to s$ HH$\pm2$) exciton (state 16, marked by a black frame), the HH$\pm2$ contribution allows a relaxation into the $p_{\text{inpl.}}\to s$ HH$\pm2$ states (states 5-8), which are of rather pure HH$\pm2$ character. Here, the index ``inpl.'' summarizes the indices $x$ or $y$. These states then relax into the dark ground states (states 1-2). In addition, the small HH$\pm1$ contribution of state 16 enables a relaxation via the rather pure $p_{\text{inpl.}}\to s$ HH$\pm1$ spin states (states 9-12) into the bright ground states (states 3-4). The ratio of the dark-to-bright spin contribution $R=W(\text{HH}\pm 2)/W(\text{HH}\pm 1)=3.8$ of the initial state 16 can be taken as a rough estimate for the ratio of the final occupation of the dark and bright ground states. The relaxation of the state $d\to s$ HH$\pm2$ (+$S\to s$ LH$\pm0$) (state 36) can be estimated in a similar way leading to $R=4.5$. Here the somehow special states 17-20 might lead to a larger disturbance of the spin state, if they are occupied during the relaxation cascade. These states contain a hole state excited in growth direction ($p_z$-state), which leads to a different mixture by valence band mixing.

Besides these rather qualitative arguments, the question for a more quantitative description of the relaxation channels emerges. To the best of our knowledge, there are still no reliable approaches to quantitatively estimate the relaxation through tens of intermediate states. However, the available approaches might give an additional hint on the final occupation of the ground states. Thus, we implemented a model based on acoustic phonon emission and polaron decay (see App.~\ref{sec:relaxapp}). This approach results in a final ratio of the occupations of the dark to bright exciton ground states of $\tilde{R}=32$ for an initial excitation of the $S\to s$ LH$\pm0$ (+$d\to s$ HH$\pm2$) exciton and a ratio of $\tilde{R}=4$ for an initial excitation of the $d\to s$ HH$\pm2$ (+$S\to s$ LH$\pm0$) exciton. Here we want to emphasize that these numbers should be treated with some care because, for example, multi-phonon processes have been neglected which might, in particular be relevant, where single-phonon processes lead to a bottleneck in the relaxation process. However, it is not expected that such extensions qualitatively change the relaxation paths. Therefore, $\tilde{R}$ provides an additional hint to the preferred final occupation of the dark exciton ground state. When comparing the two estimates $R$ and $\tilde{R}$, we find that there may be quantitative differences; however, both numbers indicate a dominant relaxation into the dark exciton ground states starting from both initial states 16 and 36 exhibiting the strong $S\to s$ LH$\pm0$ and $d\to s$ HH$\pm2$ mixing, which further supports the viability of our proposal.

\section{Dependence on the QD geometry}\label{sec:resonance}

\begin{figure}[tb]
\includegraphics[width=1.0\columnwidth]{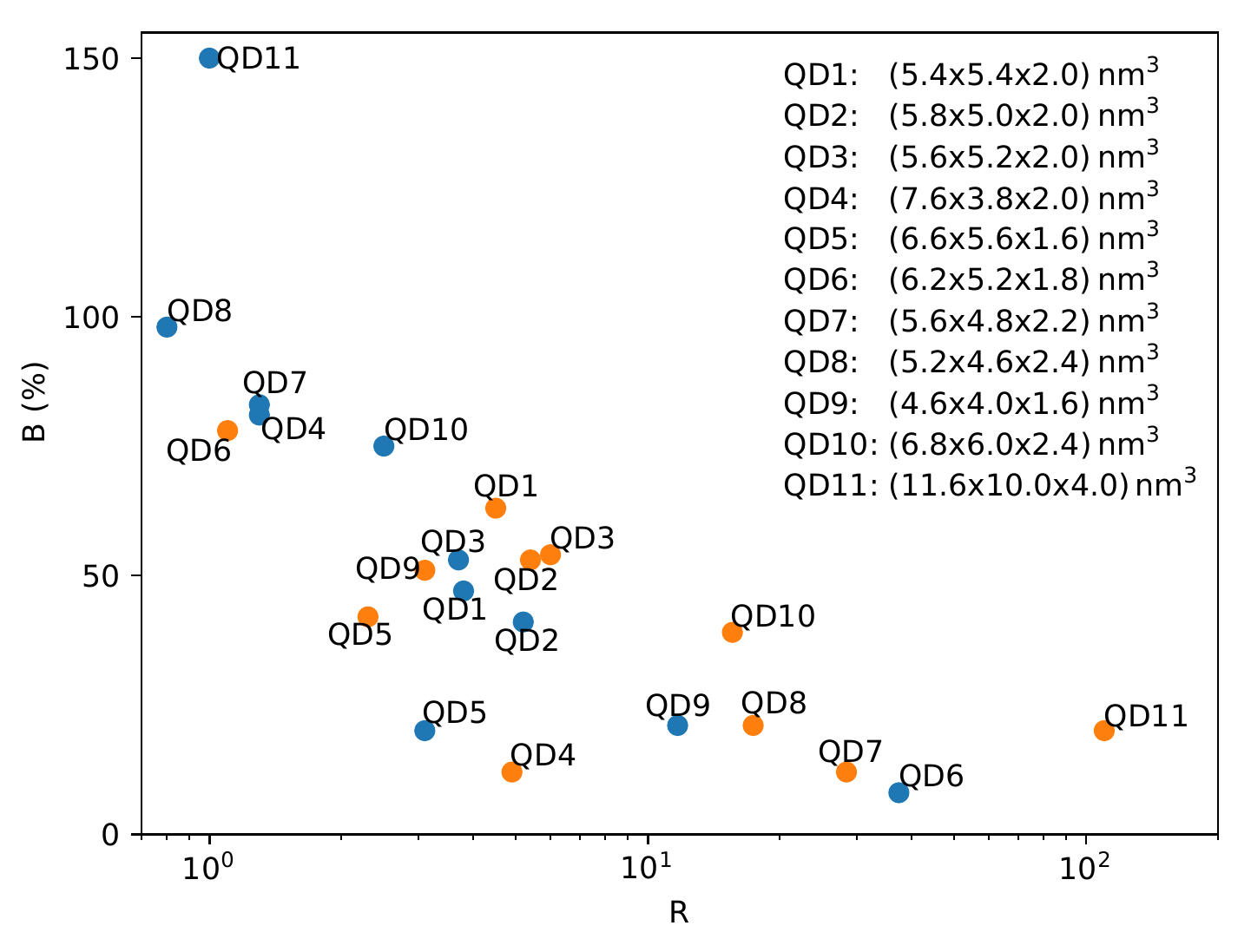}
\caption{
	\label{fig:brplot}
	Relative dark-to-bright contribution $R$ and relative brightness $B$ of the most relevant excitons (see text) excited by anti-parallel Bessel beams in different realistic QD geometries. The energetically lower/higher relevant exciton is given in blue/orange. The numerical values are also given in Tab.~\ref{tab2}.
	}
\end{figure}

In the above study, we considered a geometry where the utilized mixture between $S\to s$ LH$\pm 0$ and $d\to s$ HH$\pm2$ turned out to be maximal. The question arises whether our proposed excitation scheme crucially depends on the geometry parameters and might therefore just be relevant for very specific QDs. To rule out this possibility, we have performed a parameter study as in Ref.~\onlinecite{holtkemper2018influence} on a variety of realistic QD geometries. For each geometry, we considered two eigenstates: The one mainly consisting of $S\to s$ LH$\pm0$ spin contributions, and the brightest eigenstate mainly consisting of dark HH$\pm2$ spin contributions. In Fig.~\ref{fig:brplot} the obtained values of the relative brightness $B$ are  plotted versus the dark-to-bright ratio $R=W(\text{HH}\pm 2)/W(\text{HH}\pm 1)$ between HH$\pm2$ and HH$\pm1$ spin contribution. The energetically lower (higher) of the two considered exciton states is plotted in blue (orange).

We observe rather large variations in the brightness from $8\%$ to $150\%$ and in the dark-to-bright ratio from slightly below 1 up to about 100. Regarding the lowest values, it should be noted that the minimal observed brightness $B=8\%$ is still several orders of magnitude stronger than a direct excitation of the dark exciton \cite{schwartz2015deterministic}. Taking $R$ as a rough estmate for the final occupation of the ground states, for the lowest values $R\approx1$ we would get a $50\%$ chance to occupy the dark exciton. Thus, an excitation of these states would still have a comparable efficiency as an excitation into the wetting layer \cite{trumm2005spin}. However, here we excite a state already within the QD and with a faster and more direct relaxation into the ground state. Despite these worst case scenarios, some geometries provide states with quite appealing combinations such as $R=110$ and $B=20\%$ (QD11) or $R=15.6$ and $B=39\%$ (QD 10). 

Interestingly, we find a group of dots (QD1, QD2, QD3) for which both considered exciton states exhibit similar values for $B$ and $R$ in the intermediate range of $B\approx50\%$ and $R\approx5$. For another group (QD6, QD7, QD8, QD11) one of the two states has a rather low $R\approx1$ but a high value of the brightness $B\gtrsim80\%$ while the other state has a rather low brightness $B\approx10-20\%$ but high dark-to-bright ratios $R\gtrsim20$. In a QD from the latter class one might therefore choose between a highly efficient optical excitation with a roughly equal occupation of bright and dark excitons or a somewhat less efficient optical excitation for which more than $95\%$ of the generated occupation is expected to be in the dark state.   

An alternative way to estimate the final occupations is provided by the explicit phonon-mediated decay model introduced above and discussed in App.~\ref{sec:relaxapp}. The dark-to-bright ratio $\tilde{R}$ obtained from this model is taken as the horizontal axis of Fig.~\ref{fig:brplotnormalQD1neu} in App.~\ref{sec:relaxapp} while the vertical axis again displays the brightness $B$. When comparing the values of $R$ and $\tilde{R}$ for a specific QD there are typically some quantitative differences, as already discussed above for the exemplary QD1, however the overall shapes of Fig.~\ref{fig:brplotnormalQD1neu} and Fig.~\ref{fig:brplot} are similar supporting again the suitability of our proposal.

\section{Conclusion}\label{sec:discussion}
In conclusion, we have proposed a viable scheme to initialize the dark exciton using a single laser pulse at normal incidence. The scheme has the potential to provide much higher efficiencies than other excitation schemes. Due to long lifetimes, such dark excitons are useful for information storage in quantum technologies. Our scheme utilizes light beams with a pronounced longitudinal component tuned in resonance to certain higher exciton states. These states are characterized by a strong valence-band mixing between optically active excitons with LH$\pm0$ spin configurations and higher dark excitons with HH$\pm2$ spin configurations, which enables a large coupling to the light field and, at the same time, an ultrafast relaxation path into the dark exciton ground state. We have shown in a parameter study using a configuration interaction approach, that our findings are not specific to a certain QD geometry but appear rather generically in a wide range of QD geometries. It should be noted that no optimization of the QD properties has been performed in our study. In particular, it is known that the strength of valence-band mixing can be adjusted by strain-tuning \cite{huo2014light}, which could provide an additional tool for further optimizing the mixing between bright LH excitons and dark HH excitons. Finally we want to remark that the calculations have been performed for CdSe material parameters. However, the structure of the valence band mixing, which is the most important contribution here, is determined by the crystal structure, i.e., in the present case by the zincblende structure. Therefore we expect qualitatively similar features in QDs made from other zincblende materials.

\begin{acknowledgements}
We acknowledge support from the Open Access Publication Fund of the University of M\"unster. G. F. Q. thanks the ONRG for financial support through NICOP grant N62909-18-1-2090.
\end{acknowledgements}

\appendix

\section{Estimate of the relaxation channels}\label{sec:relaxapp}

To model the relaxation rates via acoustic phonons, we implemented a model as described in Ref.~\onlinecite{gawarecki2010phonon}, Sec. VI. Here we assume zero temperature. To account for the relaxation via LO phonons, we add a relaxation rate deduced from a polaron decay model following Ref.~\onlinecite{grange2007polaron}. To adapt the polaron decay times to our otherwise purely electronic model, we consider for each possible relaxation step just the initial and final excitonic eigenstates. Regarding these two excitonic states, we calculate the respective polaron states and use the relaxation time of the polaron closer to the respective initial excitonic state as the relaxation time between the two purely excitonic levels. The rates are then treated in a rate equation model to calculate the final occupation of dark and bright exciton ground states (the relaxation from bright to dark ground state is negligible). Such calculations are numerically quite demanding and typically just performed for single transitions (and not for the here relevant hundreds of possible single relaxations). Consequently, we reduce our basis for the calculation of the state mixtures to just slightly over 200 basis states, however, we keep the energies of the individual states as they are in the main calculation presented, e.g., in Fig.~\ref{fig:spectra} (b). We use the material parameters for CdSe: static dielectric constant $\epsilon_r=9.2$ \cite{laheld1997excitons}, density $\rho=5.81$~g/$\text{cm}^3$~\cite{madelung2012semiconductors}, lattice constant $a_{\text{lat}}=0.6077$~nm~\cite{madelung2012semiconductors}, sound velocity LA phonons $c_{\text{LA}}=3856$~m/s~\cite{madelung2012semiconductors}, sound velocity TA phonons $c_{\text{TA}}=1521$~m/s~\cite{madelung2012semiconductors}, deformation potential constant holes $D^{\text{DP}}_{h}=3$~eV~\cite{salvador2006exciton}, deformation potential constant electrons $D^{\text{DP}}_{e}=-6$~eV~\cite{salvador2006exciton}, piezoelectric constant $e_{14}=0.2$~C/$\text{m}^2$~\cite{martin1972piezoelectricity}, band center energy LO phonons $\hbar \omega_{\text{LO}}=25$~meV~\cite{schwarz2004dekker}, band edge energy LO phonons $\hbar \omega_{\text{LO}}=23$~meV, high-frequency dielectric constant $\epsilon_{\infty}=5.8$~\cite{gorska1974application}, Gr\"uneisen constant LA phonons $\gamma_{\text{LA}}=1.1$~\cite{lange2009raman}, Gr\"uneisen constant TA phonons $\gamma_{\text{TA}}=3.2$~\cite{arora1990gruneisen}.

\begin{figure}[tb]
\includegraphics[width=1.0\columnwidth]{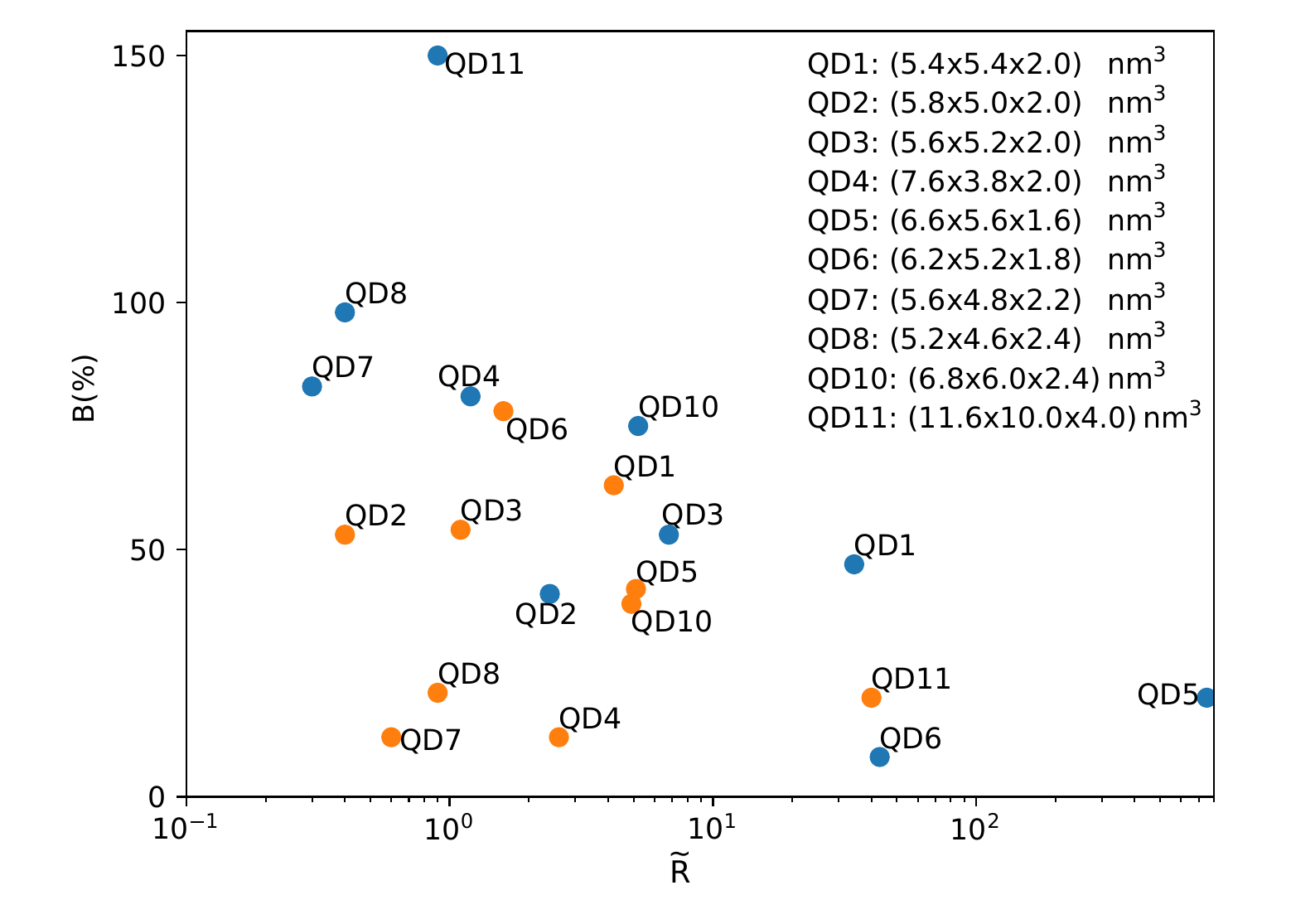}
\caption{
	\label{fig:brplotnormalQD1neu}
	Relative final occupation of dark/bright ground state $\tilde{R}$ obtained from the relaxation model and relative brightness $B$ of the same excitons as in Fig.~\ref{fig:brplot} in different realistic QD geometries. The energetically lower/higher relevant exciton is given in blue/orange. 
	}
\end{figure}

Results from these calculations are presented in Fig.~\ref{fig:brplotnormalQD1neu}, which shows in analogy to Fig.~\ref{fig:brplot} the brightness $B$ compared to the relative final occupation of dark to bright ground state $\tilde{R}$. The results support the general trend already found in the estimate via the factor $R$ (see Fig.~\ref{fig:brplot}). Differences are caused by the strong energetic selection induced by the polaron model. The energetic selection competes with the spin selection and changes the final occupation of specific states. However, there is no general trend favoring a higher final contribution of the bright or dark ground state. Considering the fact that we compare the results of completely different models, the general trend is surprisingly similar.

We want to emphasize, that these calculations should also be taken with some care and are just performed to provide another estimate on a preferred occupation of the dark exciton ground state after excitation of some suitable states. The calculations typically underestimate the total relaxation rates. They give a strong resonance-like energy dependency of the relaxations and give, e.g., for the smallest of the considered QD geometries (QD9) no relaxation at all, since the energetic level spacing is too large. This is the reason why QD9 does not appear in Fig.~\ref{fig:brplotnormalQD1neu}. Small variations in the properties of the QD might result in rather large changes of the relaxation channel if energetic resonances are met. Including, e.g., multi-phonon processes or other scattering mechanisms will provide pathways to overcome such bottlenecks and in general broaden sharp resonances. However, it is not expected that they change the general trend found here.

\section{Spin contributions for other QD geometries}\label{sec:furtherfig}

\begin{figure}[tb]
\includegraphics[width=1.0\columnwidth]{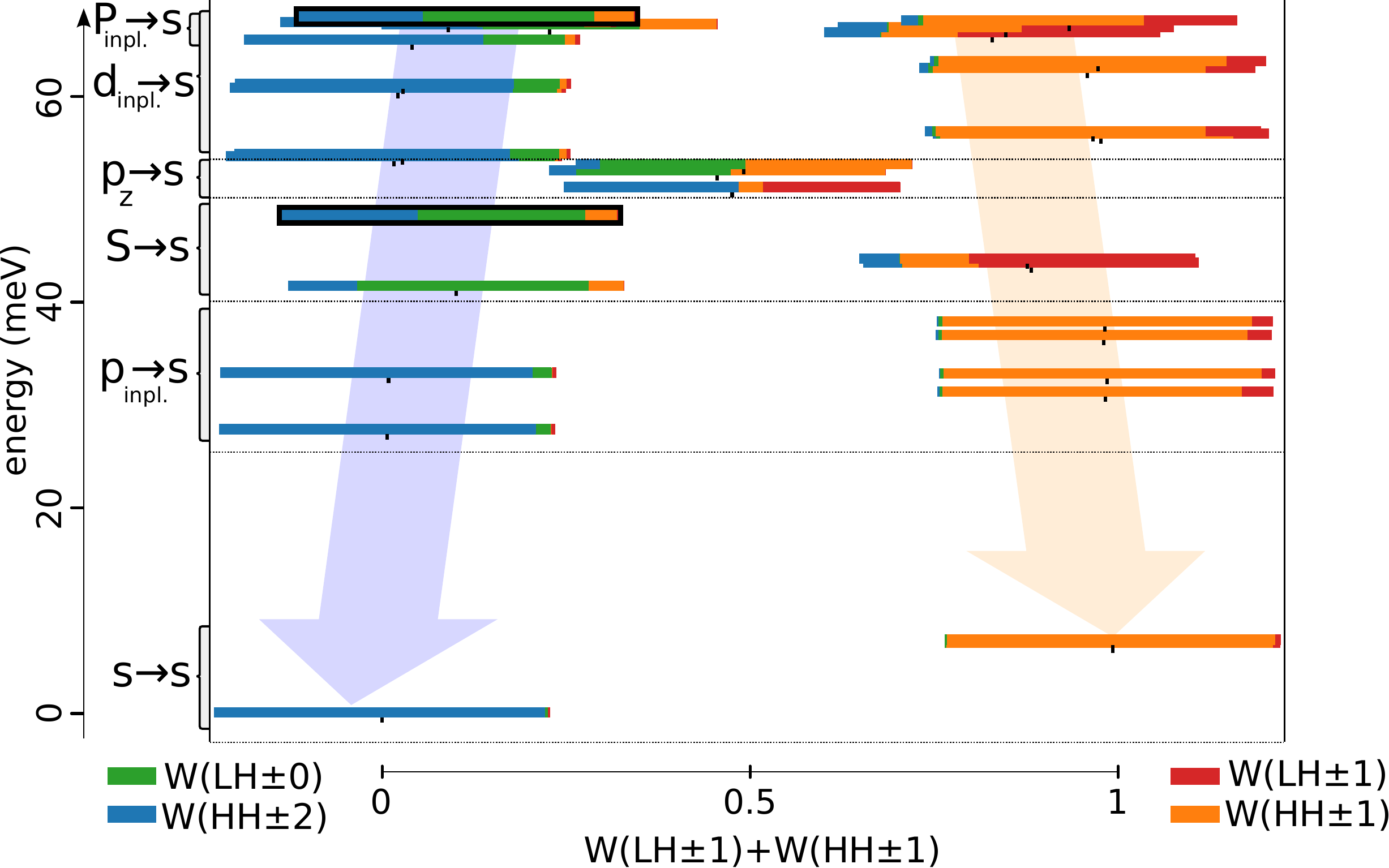}
\caption{
	\label{fig:contribrelaxin2}
	Same as Fig.~\ref{fig:contribrelaxin}, but for QD2 with a geometry of $(5.8 \times 5.0 \times 2.0)$~nm$^3$.
	}
\end{figure}

\begin{figure}[tb]
\includegraphics[width=1.0\columnwidth]{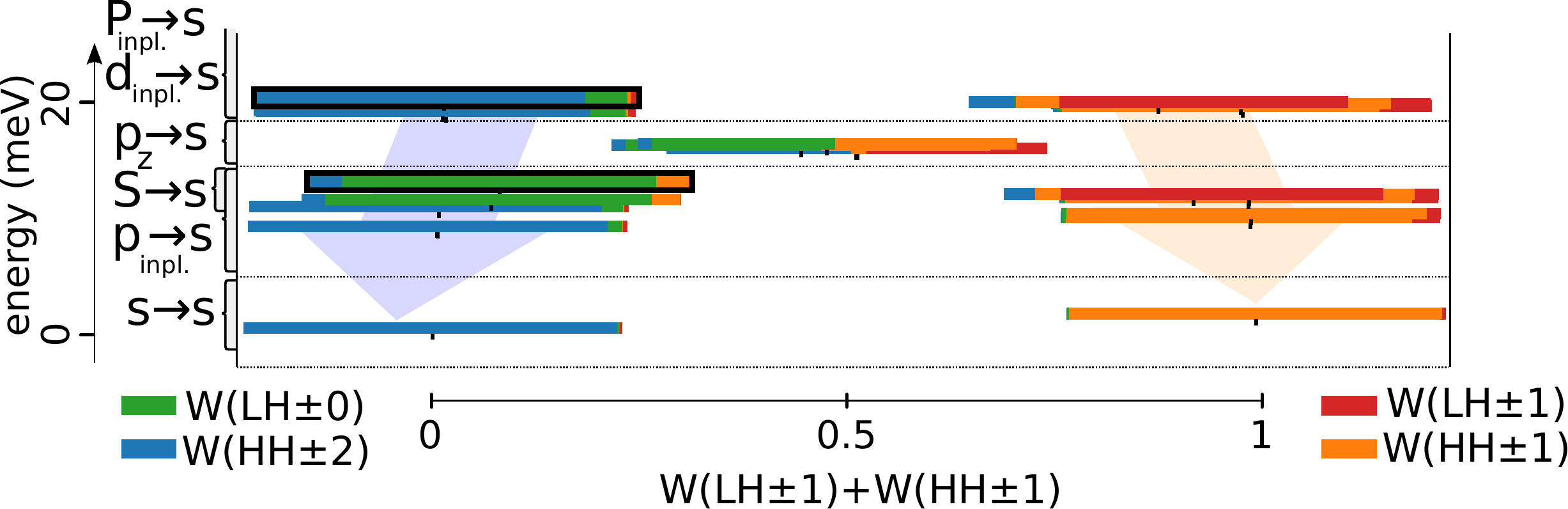}
\caption{
	\label{fig:contribrelaxin11}
	Same as Fig.~\ref{fig:contribrelaxin}, but for QD11 with a geometry of $(11.6 \times 10.0 \times 4.0)$~nm$^3$.
	}
\end{figure}

\begin{figure}[tb]
\includegraphics[width=1.0\columnwidth]{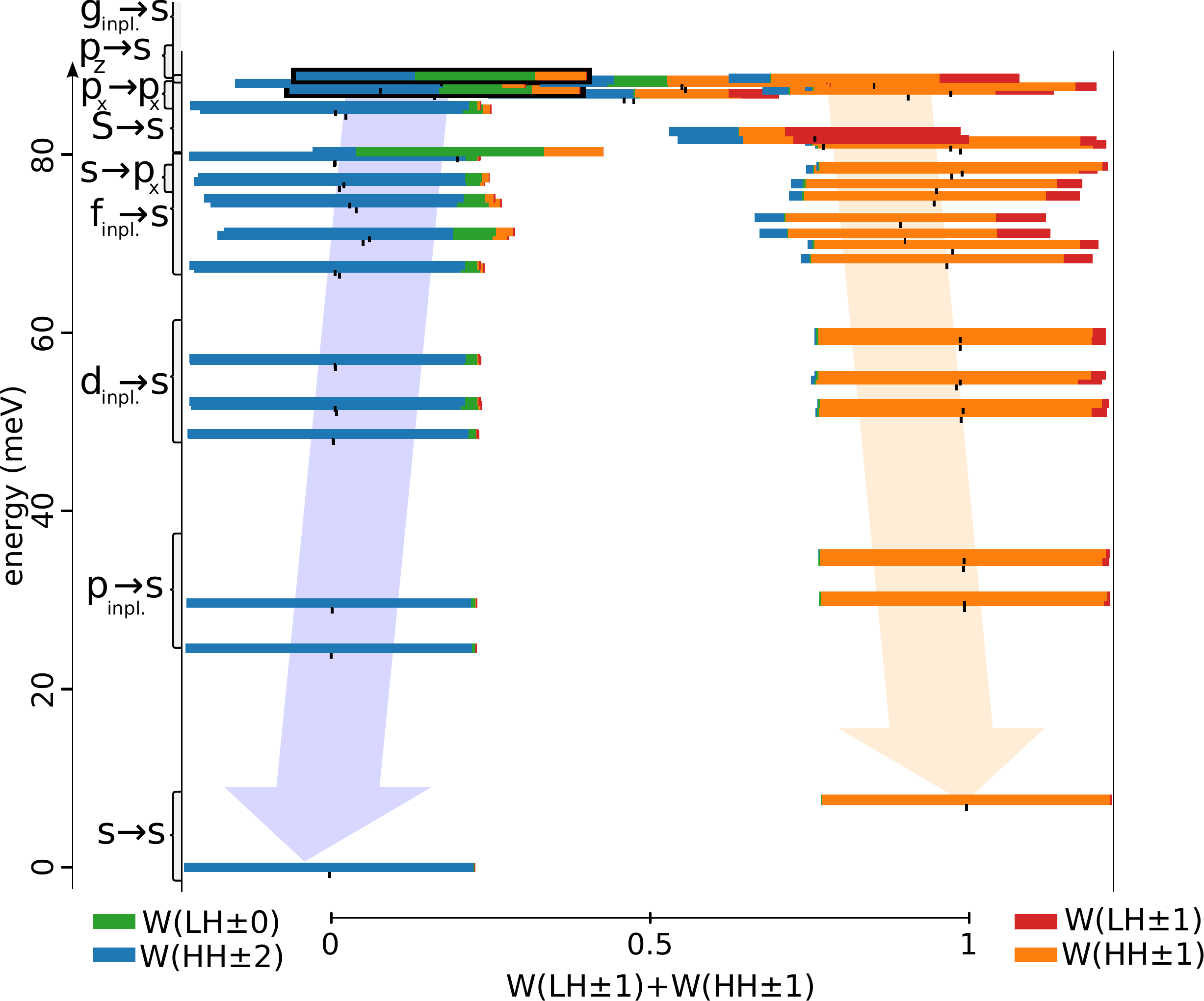}
\caption{
	\label{fig:contribrelaxin5}
	Same as Fig.~\ref{fig:contribrelaxin}, but for QD5 with a geometry of $(6.6 \times 5.6 \times 1.6)$~nm$^3$.
	}
\end{figure}

\begin{figure}[tb]
\includegraphics[width=1.0\columnwidth]{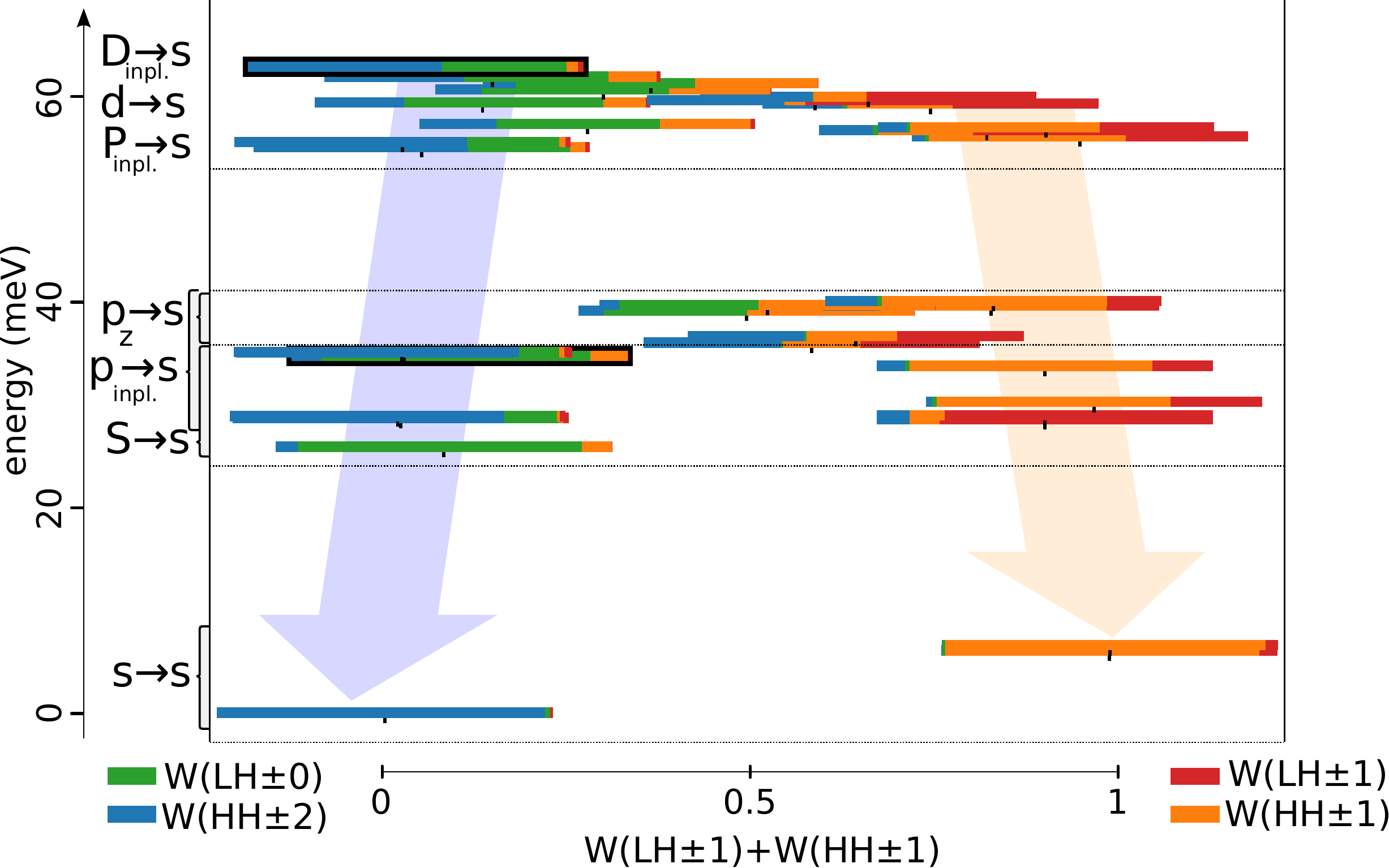}
\caption{
	\label{fig:contribrelaxin8}
	Same as Fig.~\ref{fig:contribrelaxin}, but for QD8 with a geometry of $(5.2 \times 4.6 \times 2.4)$~nm$^3$.
	}
\end{figure}

Figure \ref{fig:contribrelaxin} represents a very typical case for the spin contributions of the lower exciton states. To demonstrate this, in Figs.~\ref{fig:contribrelaxin2} - \ref{fig:contribrelaxin8} we plot the spin contributions of the lower exciton states for QD2, QD11, QD5, and QD8 in the same way as in Fig.~\ref{fig:contribrelaxin} for QD1. Note that for a better comparability the energy scale in all figures is the same. The elongation (QD2, Fig.~\ref{fig:contribrelaxin2}) and the volume (QD11, Fig.~\ref{fig:contribrelaxin11}) have nearly no impact on the overall behavior of the spin mixtures, even if for larger volumes the energy separations are strongly reduced. In contrast, the height shifts the relative position between LH and HH states. In very flat QDs (QD5, Fig.~\ref{fig:contribrelaxin5}), the LH excitons are shifted to higher energies, which simultaneously reduces the spin mixtures of the lower levels. In a very high QD (QD8, Fig.~\ref{fig:contribrelaxin8}), the LH excitons are at lower energies and there is a larger number of states with strongly mixed spin-character already at low energies.

\section{Tables}\label{sec:tables}

To provide explicit values of our results presented in Figs.~\ref{fig:anteileplot} and \ref{fig:brplot}, we present two tables listing the state mixtures of the most relevant states. The quantitatively given results are estimates within the limitations of our model.

\begin{table}[tb]
\center
 \begin{tabular}{c|c|c!{:}c!{:}c!{:}c|c|c}
 coupling &  & \multicolumn{4}{|c|}{spin contributions} &  &  \\
to light & transition & HH$\pm2$ & HH$\pm1$ & LH$\pm0$ & LH$\pm1$ & $R$ & $B$ \\
\hline
\hline
dark               & $s\to s$ & 0.99 & 0.00 & 0.01 & 0.01 & - & - \\
\hline
transv.            & $s\to s$ & 0.00 & 0.97 & 0.01 & 0.02 & 0.0    & 126\% \\
field              & $S\to s$ & 0.12 & 0.23 & 0.00 & 0.65 & 0.5    & 39\% \\
                   & $d\to s$ & 0.02 & 0.79 & 0.01 & 0.18 & 0.0    & 7\% \\
                   & $d\to s$ & 0.05 & 0.63 & 0.02 & 0.31 & 0.1    & 18\% \\
                   & $D\to s$ & 0.23 & 0.25 & 0.01 & 0.51 & 0.9    & 9\% \\
\hline
long.              & $S\to s$ & 0.39 & 0.10 & 0.51 & 0.00 & 3.8    & 47\% \\
field              & $d\to s$ & 0.48 & 0.11 & 0.40 & 0.02 & 4.5    & 63\% \\
                   & $D\to s$ & 0.24 & 0.22 & 0.54 & 0.01 & 1.1    & 33\%
 \end{tabular}
\caption{Different spin contributions of the dark exciton and the ones excited by the transverse and longitudinal fields for a QD with size $(5.4\times 5.4 \times 2.0)$~nm$^3$. $R$ is the ratio of dark to bright spin contributions and $B$ is the relative brightness of the transitions.}
 \label{tab1}
\end{table}

\begin{widetext}

\begin{table}[tb]
\center
 \begin{tabular}{c|c|c!{:}c!{:}c!{:}c|c|c|c}
QD & geometry & \multicolumn{4}{|c|}{spin contributions} & & &  \\
number & in [nm$^3$] & HH$\pm2$ & HH$\pm1$ & LH$\pm0$ & LH$\pm1$ & $B$ & $R$ & $\tilde{R}$ \\
\hline
\hline
QD1 & $5.4 \times 5.4 \times 2.0$ & 0.39 & 0.10 & 0.51 & 0.00 &  47\% & 3.8 & 34.4 \\
    &                             & 0.48 & 0.11 & 0.40 & 0.02 &  63\% & 4.5 & 4.2 \\
\hline
QD2 &$5.8 \times 5.0 \times 2.0$ & 0.45 & 0.09 & 0.46 & 0.01 &  41\% & 5.2 & 2.4  \\
    &                             & 0.52 & 0.10 & 0.37 & 0.02 &  53\% & 5.4 & 0.4 \\
\hline
QD3 & $5.6 \times 5.2 \times 2.0$ & 0.37 & 0.10 & 0.52 & 0.00 &  53\% & 3.7 & 6.8  \\
    &                             & 0.55 & 0.09 & 0.35 & 0.02 &  54\% & 6.0 & 1.1 \\
\hline
QD4 & $7.6 \times 3.8 \times 2.0$ & 0.17 & 0.13 & 0.69 & 0.00 &  81\% & 1.3 & 1.2 \\
    &                             & 0.60 & 0.12 & 0.27 & 0.01 &  12\% & 4.9 & 2.6 \\
\hline
QD5 & $6.6 \times 5.6 \times 1.6$ & 0.52 & 0.16 & 0.32 & 0.00 &  20\% & 3.1 & 751.7 \\
    &                             & 0.41 & 0.17 & 0.41 & 0.00 &  42\% & 2.3 & 5.1 \\
\hline
QD6 & $6.2 \times 5.2 \times 1.8$ & 0.83 & 0.02 & 0.14 & 0.01 &  8\%  & 37.4 & 43.0 \\
    &                             & 0.23 & 0.21 & 0.55 & 0.01 &  78\% & 1.1  & 1.6 \\
\hline
QD7 & $5.6 \times 4.8 \times 2.2$ & 0.16 & 0.12 & 0.72 & 0.00 &  83\% & 1.3  & 0.3\\
    &                             & 0.77 & 0.03 & 0.19 & 0.02 &  12\% & 28.4 & 0.6 \\
\hline
QD8 & $5.2 \times 4.6 \times 2.4$ & 0.09 & 0.11 & 0.80 & 0.00 &  98\% & 0.8  & 0.4 \\
    &                             & 0.58 & 0.03 & 0.37 & 0.02 &  21\% & 17.4 & 0.9 \\
\hline
QD9 & $4.6 \times 4.0 \times 1.6$ & 0.64 & 0.05 & 0.30 & 0.01 &  21\% & 11.7 & - \\
    &                             & 0.43 & 0.14 & 0.41 & 0.01 &  51\% & 3.1  & - \\
\hline
QD10& $6.8 \times 6.0 \times 2.4$ & 0.27 & 0.11 & 0.62 & 0.00 &  75\% & 2.5  & 5.2 \\
    &                             & 0.69 & 0.04 & 0.25 & 0.02 &  39\% & 15.6 & 4.9 \\
\hline
QD11& $11.6\times 10.0\times 4.0$ & 0.08 & 0.09 & 0.83 & 0.00 &  150\%& 1.0  & 0.9 \\
    &                             & 0.87 & 0.01 & 0.11 & 0.02 &  20\% & 110.0 & 40.0
 \end{tabular}
\caption{Most relevant (see text) exciton states excited by anti-parallel Bessel beams in different realistic QD geometries. For each state, we provide the different spin contributions, relative brightness $B$, relative spin contribution $R$ and relative final occupation of dark to bright ground states $\tilde{R}$ obtained from the relaxation model.}
 \label{tab2}
\end{table}

\end{widetext}


%

\end{document}